\documentstyle[12pt,epsfig]{article}
\newcommand{\beq}{\begin{equation}}
\newcommand{\eeq}{\end{equation}}
\newcommand{\bef}{\begin{figure}}
\newcommand{\enf}{\end{figure}}
\newcommand{\bdis}{\begin{displaymath}}
\newcommand{\edis}{\end{displaymath}}

\newcommand{\bm}[1]{\mbox{\boldmath $#1$}}
  
\title{Intermittency in the large $N$-limit of a 
       spherical shell model for turbulence}
\author{D. Pierotti}
\begin{document}

\maketitle
\centerline{Department of Physics, University of 
L' Aquila}
\centerline{Via Vetoio 1, I-67010 Coppito, L'Aquila, Italy}
\date{ }
\medskip

\begin{abstract}
A spherical shell model for turbulence, 
obtained by coupling $N$ replicas of the 
Gledzer, Okhitani and Yamada shell model,
is considered. 
Conservation of energy and 
of an helicity-like invariant is imposed in the
inviscid limit.
In the $N\rightarrow\infty$ limit
this model is analytically soluble and is remarkably similar to the
random coupling model version of shell dynamics. 
We have studied numerically the convergence of the 
scaling exponents 
toward the value predicted by Kolmogorov theory (K41). 
We have found that the rate of convergence to the K41 solution
is linear in 1/N. 
The restoring of Kolmogorov law has been related to the behaviour of the
probability distribution functions of the instantaneous scaling exponent.
\end{abstract}

PACS numbers 47.27.Jv, 47.90.+a, 05.45.+b
\newpage
\renewcommand{\baselinestretch}{2.} 

The small-scale statistics of a fluid in a regime of fully-developed
turbulence is an interesting and 
still open question. Among  the many problems which have to be faced
there is the understanding of anomalous scaling of structure functions
$S_{p}(r)$ defined as: 
\beq
S_p(r) \equiv 
< |{\bm{ v}}({\bm x}+{\bm r})-{\bm v}({\bm x})|^p> \equiv 
<|\delta_r v|^p >\sim r^{\zeta_{p}}\;\;\; r=|{\bm r}|,
\eeq
for $r$ in the inertial range, i. e. for those scales, 
much smaller than the integral scale and much larger than the viscous scales,
where inertial non-linear terms of Navier-Stokes (N-S) equation are dominant.

The phenomenological theory of Kolmogorov (K41) is
based on the hypothesis of an energy cascade 
from large to small spatial scales, local in scale size, in which
all statistical information about large scale is lost, save for the
mean energy-dissipation rate. It
leads, by simple dimensional 
arguments, to the celebrated K41 law: $\zeta_{p}^{K41}=p/3$ \cite{k41}.
This result is exact for $p=3$ and in good agreement with the experimentally 
and numerically measured exponent for $p=2$
- the energy spectrum scaling $E(k)\sim k^{-5/3}$
is very well fitted -  
but there are strong
evidences of its breakdown for larger p.

Phenomenological theories that modify the Kolmogorov approach
assigning a key role to the statistics of the spatial distribution
of energy dissipation have been introduced in the last thirty years
\cite{k62}-\cite{frish}.
These approaches are able to fit
with good accuracy the intermittent deviations from the $p/3$
law but do not have
any direct link with the dynamical 
evolution of N-S equations.
On the other hand, the goal of analytically obtaining the corrections
to K41 exponents directly from N-S equations is far from being achieved.
Until now, neither in closure theories 
nor by a renormalization group approache have 
any satisfying results been obtained 
on this problem.

A very interesting approach was proposed by Kraichnan in 1961 \cite{rcm}.
In the so-called random coupling model (RCM)  
he replaced the N-S equation
by a set of modified equations having the same basic structural properties
of N-S: quadratic non-linearity, non-linear quadratic invariants,
existence of truncated inviscid equipartition solutions. 
The velocity field is replaced by $N$ fictitious random fields 
${\bm u}^{\alpha}({\bm k},t)$ satisfying the equations:
\beq
\left(\frac{d}{dt}+\nu k^{2}\right)
u_{i}^{\alpha}({\bm k},t)
=-\frac{i}{N}k_{m}P_{ij}({\bm k})
\int_{{\bm p}+{\bm q}={\bm k}}
\Phi_{\alpha\beta\sigma}
u_{j}^{\beta}({\bm p},t)u_{m}^{\sigma}({\bm q},t)
d{\bm p}
\eeq
where $\Phi_{\alpha\beta\sigma}$ are quenched random phases
subject to suitable restrictions (see \cite{rcm}). 
The closure problem can be solved in the
$N\rightarrow\infty$ limit and gives the same equations of
direct-interaction 
approximation (DIA), a two-point closure theory introduced 
earlier by Kraichnan himself \cite{dia}. 
Recently Mou and Weichman 
\cite{mou-weic} 
introduced an alternative large-N model, the spherical model (SPM),
by taking non-random 
$\Phi_{\alpha\beta\sigma}$ coefficients and imposing additional symmetries
on the system. 
Their idea is basically that of
1/N expansion of critical phenomena: one
generalizes the model under consideration to one with an higher 
symmetry $G_{N}$
(in standard application, for example, 
one considers the rotation group $O(N)$, with physical values $N=1,2,3$);
the $N\rightarrow\infty$ limit  
is analytically soluble and a systematic expansion in powers of 
$1/N$ may be developed \cite{ma}. 
Mou and Weichman, following the work of Amit and Roginsky \cite{amit},
chose the group $G_{N}=O(3)$ for all $N$ but allowed 
the dimension of the representation to diverge with $N$.
The remarkable point is that
the SPM and the RCM are similar: in both the infinite $N$ 
limit leads to the DIA 
equations \cite{eyink}.  
Both models, then, are plagued with the same problem:  
DIA equations violate the basic physical principle of
galilean invariance and give, consequently,
the wrong energy spectrum scaling,
$E(k)\sim k^{-3/2}$.
The correct -5/3 exponent has been recovered, in fact, in a 
Lagrangian modification of DIA in which galilean invariance is restored
(for an overview about these theories see \cite{lesieur}).
So it would be ill advised to try to get anomalous corrections of scaling
exponents starting from the wrong zeroth order and taking successive orders
in which the same problems occur.
However, the basic idea of getting the intermittent corrections
as asymptotic expansions in powers of $1/\sqrt{N}$ 
\cite{mou-weic}-\cite{eyink}, i. e.
\begin{equation}
\delta\zeta_{p}\equiv\zeta_{p}^{K41}-\zeta_{p}
=\sum_{k> 1} \frac{c_{k}^{(p)}}{N^{k/2}}
\label{eq:expans}
\end{equation}
can be tested on simple models in which these inconsistencies do not
show up like, for example, the shell models, 
as Eyink has
recently proposed \cite{eyink}. 

Shell models are dynamical systems 
in which, exploiting the idea that active
degrees of freedom in a turbulent flow
are hierarchically organized, 
a reduced number of variables (one or two) are considered for 
each shell of wavevectors with modulus $k\in(k_{n}, k_{n+1}]$, where
$k_{n}=k_{0}2^{n}$.  
Among shell models, the so-called GOY (from Gledzer, Ohkitani and Yamada)
\cite{G}-\cite{OY} is of considerable recent interest.  
The model is defined by the following ODE's:
\begin{equation}
\frac{d}{dt}u_{n} = -\nu k_{n}^{2} u_{n} + i k_{n} 
\left[ u_{n+1} u_{n+2}
-\frac{\varepsilon}{2} u_{n-1}u_{n+1}
-\frac{(1-\varepsilon)}{4} u_{n-1} u_{n-2} \right]^{\ast} 
+f_{n}\delta_{n,1}
\label{eq:shellmodel}
\end{equation}
with $n=1,...,N_{shells}$ and
boundary conditions with null
$u_{-1}$, $u_{0}$, $u_{N_{shells}+1}$ and $u_{N_{shells}+2}$.
The coefficients of the non-linear terms have been chosen 
in order to keep
the energy constant in the inviscid unforced case;  
besides, for $\varepsilon =1/2$, one has an inviscid invariant
$H=\sum_{n}(-1)^{n}k_{n}|u_{n}|^{2}$ \cite{benzi} that is a 
sort of ``shell model helicity'' similar to the helicity
$H=\int ({\bm k}\times {\bm v}({\bm k}) 
)\cdot
{\bm v}({\bm k}) d{\bm k}$ conserved in 
the 3d Euler equations. 
Considering
that $u_n$ should describe velocity fluctuations
at scale $r=1/k_n$ the structure functions are defined by:
\beq
S_{p}^{n} \equiv < |u_n|^p> 
\label{eq:zetap_sm}
\eeq
and show, in the inertial range, a scaling behaviour: 
$S_{p}^{n}\sim k_n^{-\zeta_{p}}$.
This model 
has the remarkable property of reproducing quantitatively
the exponents $\zeta_{p}$ experimentally measured when 
$\varepsilon =1/2$ \cite{JPV}-\cite{pisa}.
 
In \cite{eyink} Eyink defined and studied the infinite $N$ limit 
of the following
complex spherical shell models (SSM):
\beq
\left(\frac{d}{dt}+\nu k_{n}^{2}\right)u_{n}^{\alpha}=
\sum_{ml,\beta\gamma}A_{nml}W_{N}^{\alpha\beta\gamma}
(u_{m}^{\beta}u_{l}^{\gamma})^{*}+f_{n}^{\alpha}
\eeq
with $A_{nml}$ local couplings, i.e. vanishing outside a finite range of
neighbors and $W_{N}^{\alpha\beta\gamma}$ given by (\ref{W_N}).
This kind of models reduce to real shell model when $N=1$ and the 
DIA equations obtained in the infinite $N$ limit have stationary solutions
with K41 scaling when a fixed input of energy by the external force 
is imposed.  

We considered a slightly modified version that gives the original GOY model
at $N=1$. 
We generalized the dynamical equations for the 
real and imaginary parts of the complex variable $u_{n}=x_{n}+iy_{n}$ 
to the following dynamical system:

\begin{eqnarray}
\left( \frac{d}{dt}+ \nu k_{n}^{2} \right) x_{n}^{\alpha}
 & = & k_{n}W_{N}^{\alpha\beta\gamma}
     [ x_{n+1}^{\beta}y_{n+2}^{\gamma}
      +y_{n+1}^{\beta}x_{n+2}^{\gamma}
    -\frac{\varepsilon}{2}
     (x_{n-1}^{\beta}y_{n+1}^{\gamma}
     +y_{n-1}^{\beta}x_{n+1}^{\gamma}) \nonumber  \\
 & - & \frac{(1-\varepsilon)}{4}
	( x_{n-1}^{\beta} y_{n-2}^{\gamma}
      +y_{n-1}^{\beta} x_{n-2}^{\gamma})]^{*}
      +f^{\alpha}\delta_{n1}                   \\ 
\left( \frac{d}{dt}+ \nu k_{n}^{2} \right) y_{n}^{\alpha}
 & = & k_{n}W_{N}^{\alpha\beta\gamma}
     [ x_{n+1}^{\beta}x_{n+2}^{\gamma}
      -y_{n+1}^{\beta}y_{n+2}^{\gamma}
    - \frac{\varepsilon}{2}
     (x_{n-1}^{\beta}x_{n+1}^{\gamma}
     -y_{n-1}^{\beta}y_{n+1}^{\gamma}) \nonumber  \\
 & - & \frac{(1-\varepsilon)}{4}
	( x_{n-1}^{\beta} x_{n-2}^{\gamma}
      -y_{n-1}^{\beta} y_{n-2}^{\gamma})]^{*}
      +f^{\alpha}\delta_{n1} 
\end{eqnarray} 
where
sums over repeated indices are implied; $\alpha =-J$,...,$J$;
$N=2J+1$ and 
$n=1,...,N_{shells}$. 
To get the original model in the limit $J\rightarrow 0$
we have taken $x_{n}^{\alpha}$ and $y_{n}^{\alpha}$ complex variables
satisfying
$x_{n}^{-\alpha}=(-1)^{\alpha}(x_{n}^{\alpha})^{*}$ and
$y_{n}^{-\alpha}=(-1)^{\alpha}(y_{n}^{\alpha})^{*}$ and
$f^{-\alpha}=(-1)^{\alpha}(f^{\alpha})^{*}$. 
So there are  
$2N=2(2J+1)$ real degrees of freedom per each shell.
Following \cite{eyink} 
we imposed the invariance of the system under the transformation: 
\begin{equation}
{\bm v}_{n}^{\alpha}\rightarrow 
D_{\alpha \beta}^{N}(U){\bm v}_{n}^{\beta},
\end{equation}
where the
$D_{\alpha \beta}^{N}(U)$ are the Wigner $D^{J}$ matrices of quantum mechanics
(N-dimensional irreducible
representations of the group $SU(2)$)
and ${\bm v}_{n}^{\alpha}=(x_{n}^{\alpha},y_{n}^{\alpha})$.
So one must take:
\begin{equation}
W_{N}^{\alpha\beta\gamma}=(-1)^{\alpha}<JJJ-\alpha|JJ\beta\gamma>
\label{W_N}
\end{equation}
which are the Clebsh-Gordan coefficients.
The properties of these coefficients, together with the fact that
we considered only even J's, guarantee:
(i) invariance of the system under any permutation of the replica indices;
(ii) conservation of volume in phase-space; (iii) conservation, 
in the inviscid
unforced case,
of energy:
\begin{equation}
E=\frac{1}{2}\sum_{n=1}^{N}\sum_{\alpha=-J}^{+J}
|x_{n}^{\alpha}|^{2}+|y_{n}^{\alpha}|^{2}
\end{equation}
and of shell-model helicity:
\begin{equation}
H=\frac{1}{2}\sum_{n=1}^{N}\sum_{\alpha=-J}^{+J}(-1)^{n}k_{n}
[|x_{n}^{\alpha}|^{2}+|y_{n}^{\alpha}|^{2}].
\end{equation}
Let us stress at this point that the choice of considering the SSM rather
then a shell RCM
has been 
dictated by the considerable advantage that the former 
present in a numerical analysis, avoiding the averages on the random 
coupling coefficients.

We analyzed the scaling properties of this model at varying the number of
fields ${\bm v}^{\alpha}$ (up to a maximum number of 21 replicas).
As the moments of the PDF's of the velocity variables are affected by period
three oscillations (responsible for large uncertainties in scaling
exponents) and as quite a large precision is needed to calculate deviations
from K41 solution we considered the scaling properties of the following
suitable triple products:  
\begin{equation}
\Pi_{n}(N)=-{\cal Re}
\left(\Delta_{n+1}(N)+\frac{1-\varepsilon}{2}\Delta_{n}(N)\right)
\end{equation} 
where:
\begin{eqnarray}
\Delta_{n}(N) & = &\sum_{\alpha\beta\gamma}W_{N}^{\alpha\beta\gamma}
                \left[ x_{n}^{\alpha}
		\left(x_{n+1}^{\beta}y_{n-1}^{\gamma}+
		      y_{n+1}^{\beta}x_{n-1}^{\gamma}\right)
		\right.\nonumber \\
           &  & \left.
		+y_{n}^{\alpha}
		\left(x_{n+1}^{\beta}x_{n-1}^{\gamma}-
		      y_{n+1}^{\beta}y_{n-1}^{\gamma}\right)
	        \right]     
\end{eqnarray} 
The variables $\Pi_{n}$ have a physical relevance, 
$\Pi_{n}k_{n}$ 
representing the energy flux from the scale n to larger wavenumber scales,
and are not plagued with periodic oscillations (see fig. 1).
We make use of a relation analogous 
to the structure equation of Kolomogorov
\cite{monin} as in 
\cite{benzi} and so the following scaling law
is expected:
\begin{equation}
\Sigma_{p}^{n}(N)=\langle|\Pi_{n}(N)|^{p/3}\rangle\sim
 k_{n}^{-\zeta_{p}(N)}
\end{equation} 
In figure 2 we plotted the behaviour of 
$\delta\zeta_{p}(N)=\zeta_{p}^{K41}-\zeta_{p}(N)$ 
as function of 1/N for different
values of $p$. The corrections to the K41 solution tend to zero when 
$N\rightarrow\infty$ and the rate of convergence is linear in $1/N$.
It should then be possible to develop an asymptotic expansion for the
$\delta\zeta_{p}$'s of the form (\ref{eq:expans}) as Eyink suggested 
for the shell models in
\cite{eyink}.

In our numerical simulations we used  
$N_{shells}=15$ and 19, with $\nu=10^{-5}$ and $10^{-6}$,
$f^{\alpha}=5\;10^{-3}(1+i)$, $k_{0}=6.25\;10^{-2}$ and a number of replicas
up to 21. Equations have been integrated by a slaved 
Adams-Bashforth algorithm and temporal averages are over several thousands
of eddy turn-over times of the first shell. 

Beyond these numerical results an interesting physical problem is to 
understand the dynamical mechanism by which K41 is restored.
As Kraichnan pointed out in \cite{Kraic}, the K41 law is not ruled out just 
because energy-dissipation fluctuates: it could be {\it a priori} possible
if there were spacewise diffusion effects of sufficient strength to suppress
fluctuations of energy transfer in the inertial range. 
We suppose that a similar mechanism is present in the SSM, although it is
not completely clear to us to what extent the similarity can be pushed.
In the GOY model the anomalous scaling law is accompanied by an intermittent
behaviour of the energy transfer in the inertial range with very large
fluctuations. A quantity that measure the local (in time) singularity is
the instantaneous
scaling exponent for fluxes, defined as:
\begin{equation}
h(t)=\frac{1}{n_{max}-n_{min}+1}\sum_{n_{min}}^{n_{max}}
\ln_{2}\left|\frac{\Pi_{n}}{\Pi_{n+1}}\right|
\label{istant}
\end{equation}
where $n_{min}$ and $n_{max}$ delimit the inertial range.
In a laminar regime one has $h=3$, $h=1$ corresponds to a K41
scaling while a value of $h<1$ represents a more singular velocity field. 
Such a value of $h$ is realized during the fast energy bursts
\cite{lima}.
We measured h(t) and calculated its PDF for different values of N.
We observed that increasing the number of coupled replicas the peak
of the distribution gets sharper and the minimum value of h increases, 
indicating that the field is getting less singular in the inertial range
(see fig. 3). This result could be due to possible exchanges of energy
between the increasing number of degrees of freedom at the same scale.

To conclude we would like to stress that the SSM is the only
model we know 
in which it has been shown that there is approach to the K41 solution
by tuning a parameter (the number of replicas) starting from an intermittent
solution. Let us remark, anyway,
that in a different contest, that of passive scalar advection,
two groups have been succeded in calculate anomalous scaling exponents  
by expansion around limiting gaussian cases \cite{Gaw,Cher}. 
At now nothing similar has been done with the N-S equation
also if it is believed that an infinite-dimension limit should 
converge to K41. Beyond the possibility of getting an asymptotic expansion
for the anomalous corrections of the form (\ref{eq:expans}) we believe that
further work in this direction should deepen the comprehension of 
intermittency in real turbulence.   

\section*{Acknowledgment}
I would like to remember G. Paladin whose enthusiasm and support
have been very important for this work and for me.
I wish to thank L. Biferale and A. Vulpiani for important discussions
on this work. 
Moreover I would like to thank G.L. Eyink and
G. Parisi for precious suggestions.

\newpage
\centerline{FIGURE CAPTIONS}

\begin{itemize}

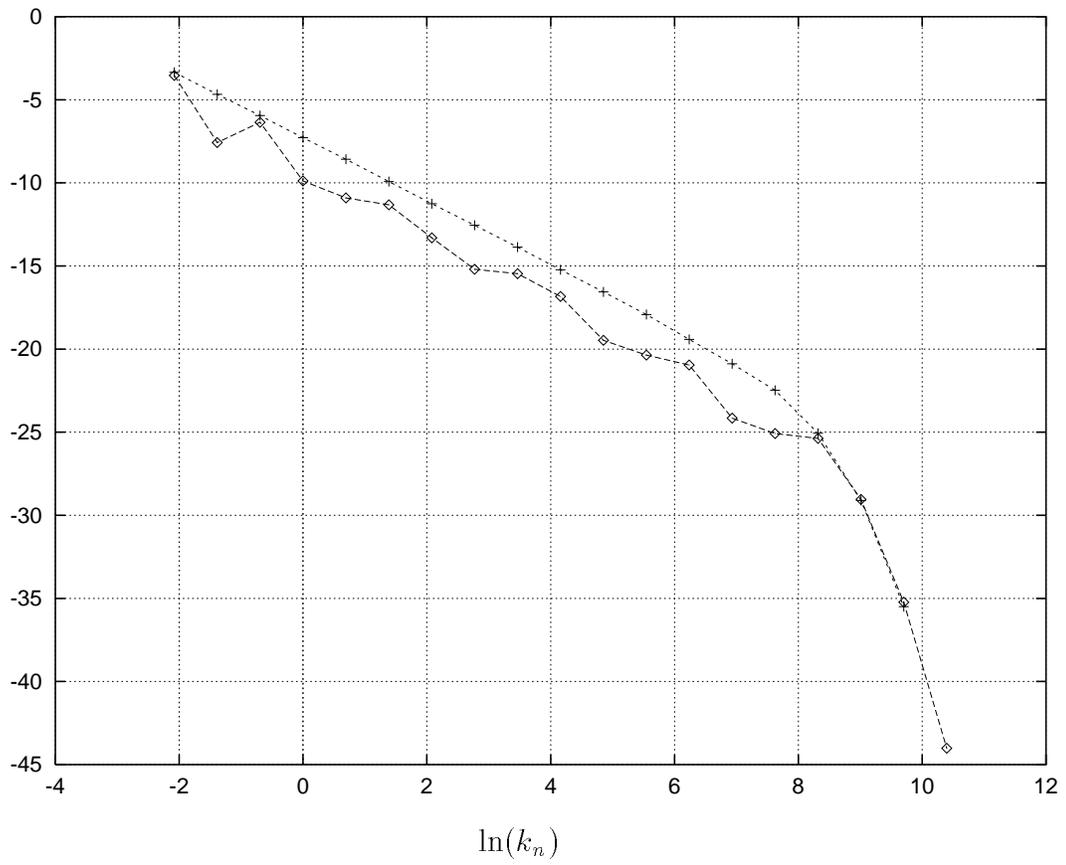
\item FIGURE 1:\\
Structure function $\Sigma_{p=6}^{n}(N)$ vs $k_{n}$ in a log-log plot
for $N=5$ and $N_{shells}=19$ (crosses with line). We plotted also the
scaling behaviour of the 6-th moment
$\left\langle|u_{n}^{\alpha}|^{p=6}\right\rangle$
for $\alpha=1$ (diamonds).

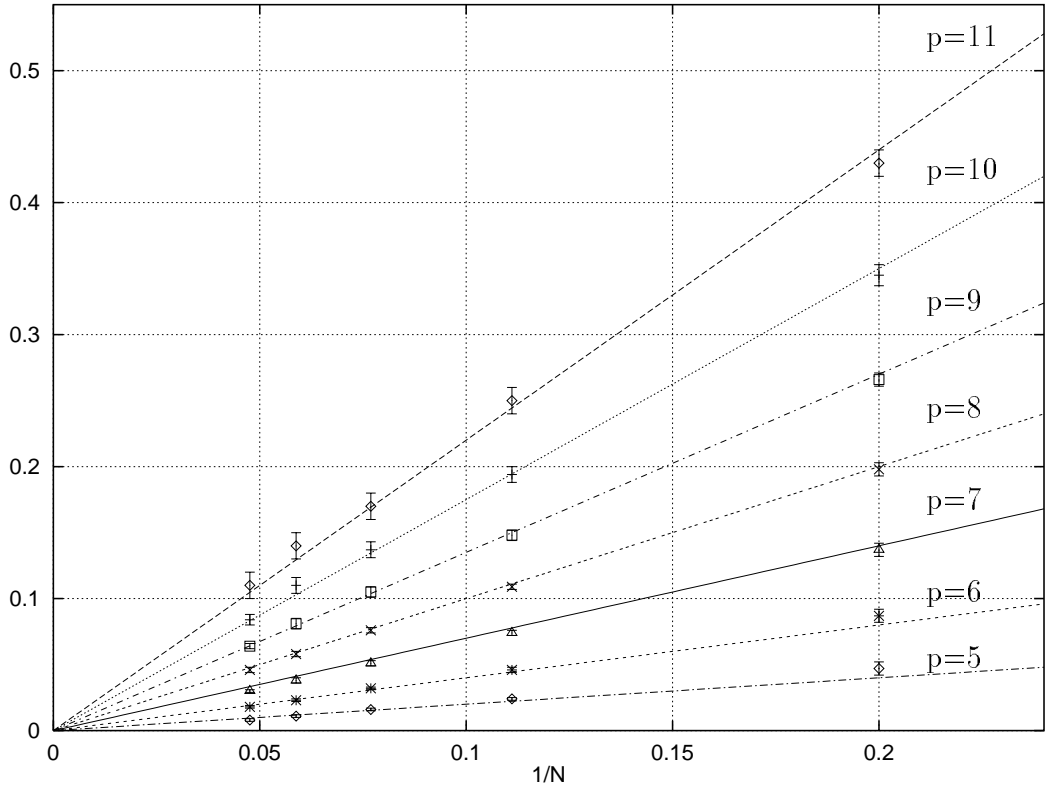
\item FIGURE 2:\\
Anomalous corrections $\delta\zeta_{p}(N)$ versus 1/N for $p=5,...,11$. 
The error bars have been obtained by a least square fit.  
Note that the origin is an analytically calculated point.

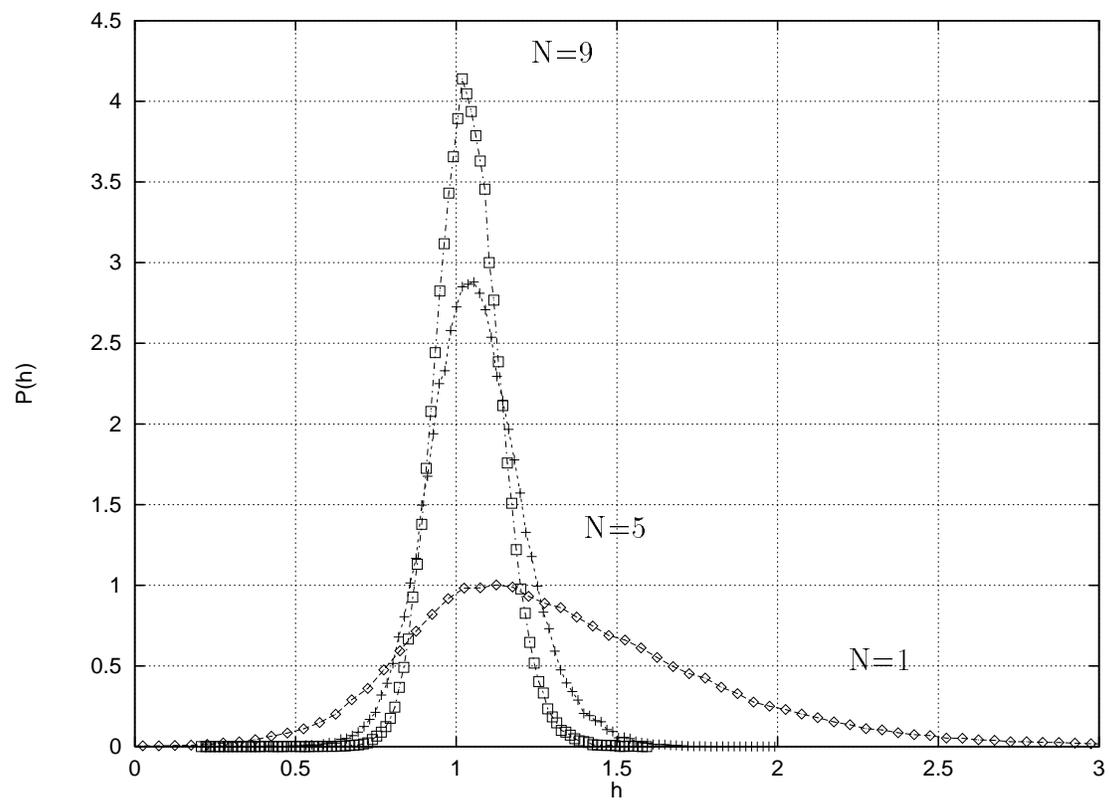
\item FIGURE 3:\\
Probability distribution function of the instantaneous scaling exponents
h(t) of equation (\ref{istant})for different values of N.
\end{itemize}

\end{document}